\documentclass[10pt,journal,twocolumns]{IEEEtran}
\usepackage{subfigure}
\usepackage{amssymb}
\usepackage{amsthm}
\usepackage{amsmath}
\usepackage{mathrsfs}
\usepackage{hyperref}
\usepackage{graphicx}
\usepackage{colortbl,dcolumn}
\usepackage{booktabs}
\usepackage{xcolor}
\usepackage{cite}
\usepackage{threeparttable}

\usepackage{algorithm}
\usepackage{algorithmic}

\usepackage{cases}

\usepackage{stmaryrd}

\newtheorem{assumption}{Assumption}
\newtheorem{definition}{Definition}
\newtheorem{lemma}{Lemma}
\newtheorem{theorem}{Theorem}

\newtheorem{corollary}{Corollary}
\newtheorem{remark}{Remark}
\newtheorem{problem}{Problem}

\newcommand{\defref}[1]{Definition~\ref{#1}}
\newcommand{\lemref}[1]{Lemma~\ref{#1}}
\newcommand{\thmref}[1]{Theorem~\ref{#1}}

\newcommand{\corref}[1]{Corollary~\ref{#1}}
\newcommand{\figref}[1]{Fig.~\ref{#1}}

\newcommand{\secref}[1]{Section~\ref{#1}}
\newcommand{\apxref}[1]{Appendix~\ref{#1}}
\newcommand{\probref}[1]{Problem~\ref{#1}}

\newcommand{\rekref}[1]{Remark~\ref{#1}}
\newcommand{\algref}[1]{Algorithm~\ref{#1}}

\usepackage{mathtools} 

\usepackage{enumerate}

\usepackage{geometry}

\ifCLASSOPTIONcompsoc
\else
\fi
\ifCLASSINFOpdf
\else
\fi

\graphicspath{{pdfs/},{images/}}

\begin{document}
%
\title{Existence and Completeness of Bounded Disturbance Observers: A Set-Membership Viewpoint}
%
%
%
%

\author{Yudong~Li,~Yirui~Cong,~Jiuxiang~Dong
\IEEEcompsocitemizethanks{\IEEEcompsocthanksitem Y.~Li and J.~Dong are with the State Key Laboratory of Synthetical Automation of Process Industries, Northeastern University, China, also with the College of Information Science and Engineering, Northeastern University, China, and also with the Key Laboratory of Vibration and Control of Aero-Propulsion Systems Ministry of Education of China, Northeastern University, China (email: 2210325@stu.neu.edu.cn; dongjiuxiang@ise.neu.edu.cn).\protect\\
\indent Y.~Cong is with the College of Intelligence Science and Technology, National University of Defense Technology, China (congyirui11@nudt.edu.cn).
%
}
}
\IEEEtitleabstractindextext{%
\begin{abstract}

This paper investigates the boundedness of the Disturbance Observer (DO) for linear discrete-time systems.
In contrast to previous studies that focus on analyzing and/or designing observer gains, our analysis and synthesis approach is based on a set-membership viewpoint.
From this viewpoint, a necessary and sufficient existence condition of bounded DOs is first established, which can be easily verified.
Furthermore, a set-membership filter-based DO is developed, and its completeness is proved;
thus, our proposed DO is bounded if and only if bounded DOs exist.
%
We also prove that the proposed DO has the capability to achieve the worst-case optimality, which can provide a benchmark for the design of DOs.
Finally, numerical simulations are performed to corroborate the effectiveness of the theoretical results.
    
\end{abstract}

}

\maketitle

\IEEEdisplaynontitleabstractindextext

%
\IEEEpeerreviewmaketitle

\section{Introduction}\label{sec:Introduction}
\subsection{Motivation and Related Work}\label{sec:Motivation and Related Work}
Disturbance observer (DO) is an effective method to estimate external disturbances and unmodeled system dynamics.
It is widely applied to modern control theory~\cite{chen2015disturbance} and has a significant impact on disturbance attenuation~\cite{Wang2016}, fault-detection~\cite{8809884}, robust control~\cite{kurkccu2018disturbance}, cyber-physical systems~\cite{10124355}, etc.
%
%
When designing a DO, one should guarantee the boundedness of the estimation error.
%
Otherwise, the estimated disturbance can hardly provide useful information on the actual disturbance.
%

%
%
For continuous-time systems, the DOs are capable of capturing the actual disturbance, i.e., the estimation error can not only be bounded, but also converge to zero.
%
%
For example,~\cite{lan2020asymptotic} provided an observer framework that can realize asymptotic estimation of state, fault, and disturbance estimation under a Linear Matrix Inequality (LMI) condition;
in~\cite{rabiee2019continuous}, an adaptive sliding mode disturbance observer was designed to realize the disturbance estimation in finite time.
%

For discrete-time systems, such convergence is not preserved in general, due to the influence of the sampling period.\footnote{For discrete-time systems, the disturbance uncertainty is accumulated over each time step, which cannot be eliminated instantly since the measurement equation usually does not contain the disturbance.
But for continuous-time systems, they are equivalent to having infinitesimal sampling periods; thus, the uncertainty is not accumulated and can be eventually eliminated.}
%
%
%
For instance,~\cite{chang2006applying} designed a proportional integral observer; the estimation error and the accumulated disturbance uncertainty (over each time step) had the same order of $O(h^2)$, where $h$ is the sampling period.
Similarly,~\cite{Kim2013} showed that the estimation error and the accumulated disturbance uncertainty shared the same order of $O(h)$ for the proposed DO.
%

To determine whether a DO is bounded (i.e., with bounded estimation error), it is essential to identify the conditions under which the boundedness can be achieved.
%
%
In the literature, seeking boundedness conditions can be classified into two main types, based on different classes of DOs:
\begin{itemize}
    \item   \textbf{Joint state-disturbance estimation class:}
    Since a disturbance can be viewed as a part of the augmented state, the disturbance is usually estimated jointly with the system state~\cite{Corless1998}.
    Recent studies~\cite{alenezi2021simultaneous} and~\cite{9872111} provided joint estimation methods with corresponding boundedness conditions.
    %
    %
    More specifically, in~\cite{alenezi2021simultaneous}, the state and the disturbance were simultaneously estimated based on a proposed unknown-input-observer architecture;
    it is shown that the estimation error is bounded when the system is detectable and satisfies an easy-to-verify LMI condition.
    %
    In~\cite{9872111}, an iterative proportional-integral observer was first designed to give an interval estimation for the state and the disturbance, with the observability and a provided LMI condition to guarantee the boundedness of the estimation error.
    Note that the joint estimation can be regarded as a full-order DO, where the whole system state is involved.
    %
    
    \item   \textbf{State Functional Observer (SFO) class:}
    With the SFO theory~\cite{darouach2000existence}, it is possible to estimate the disturbance only using part of the system state;
    thus, the boundedness condition is generally less conservative than that of the joint estimation which requires the full state.
    %
    %
    In~\cite{Kim2013}, the authors improved the results in their early work~\cite{5457987} and proposed an SFO-based DO with the minimal order;
    the established existence condition of bounded DOs was closely associated with a static output feedback problem.
    %
    %
    Then,~\cite{Su2018} further developed the existence condition in~\cite{Kim2013}, and an easily checked necessary and sufficient existence condition was derived, given any state functional matrix related to estimating the disturbance.
    In~\cite{8732471}, an iterative procedure for input/disturbance functional observer was proposed, where the necessary and sufficient existence condition was provided.
    Nevertheless, within the SFO class, the existence condition of bounded DOs relies on the prerequisite of carefully designing a state functional matrix whose existence remains an open problem.
    %
\end{itemize}

To the best of our knowledge, the necessary and sufficient condition under which bounded DOs exist is still unknown;
as a result, the ability of the existing DOs to ensure a bounded estimation error is inadequately understood.
Therefore, it is crucial to establish the necessary and sufficient existence condition of bounded DOs and develop a complete algorithm to guarantee boundedness whenever the condition is satisfied.


%
%

%
%

\subsection{Our Contributions}\label{sec:Our Contributions}

From a set-membership viewpoint, this article first establishes a necessary and sufficient existence condition of bounded DOs for linear discrete-time systems, and designs a complete algorithm to achieve the boundedness.
%
%
The main contributions are as follows:
\begin{itemize}
\item   By introducing the uncertain variable~\cite{6415998}, the bounded DO is rigorously formulated in a set-membership manner.
    Then, we put forward an explicit necessary and sufficient existence condition of bounded DOs.
    With this condition, one can easily determine whether bounded DOs exist.

\item   We propose an SMF-based DO, and its completeness is proved.
    Consequently, our proposed DO is bounded if and only if bounded DOs exist (easily to check using the existence condition).
    Furthermore, we prove that the proposed DO is able to achieve the worst-case optimality, thereby providing a benchmark for the design of DOs.

%
\end{itemize}
\subsection{Notation and Preliminaries}\label{sec_Notation and Preliminaries}
Throughout this paper, we use $\left\|  \cdot  \right\|$ to represent the Euclidean norm (of a vector) or the spectral norm (of a matrix),  and ${\left\|  \cdot  \right\|_\infty }$ to represent the infinite norm.
For a sample space $\Omega$, a measurable function ${\mathbf{x}}:\Omega  \to \mathcal{X}$ from sample space $\Omega$ to measurable set $\mathcal{X}$, expressed by upright letters, is called an uncertain variable~\cite{6415998,cong2021rethinking}, with its range $\llbracket\mathbf{x}\rrbracket$ defined by:
\begin{equation}\label{eq_uncertain_varibale}
\llbracket\mathbf{x}\rrbracket := \left\{\mathbf{x}(\omega)\colon \omega \in \Omega\right\}.
\end{equation}
$D(\mathcal{S}{\text{) = su}}{{\text{p}}_{s,s' \in {\mathcal{S}_k}}}\left\| {s - s'} \right\|$ stands for the diameter of $\mathcal{S}$. 
For multiple uncertain variables with consecutive indices, we
define ${{\mathbf{x}}_{{k_1}:{k_2}}} = {{\mathbf{x}}_{{k_1}}}, \ldots ,{{\mathbf{x}}_{{k_2}}}$. 
${I_{n \times m}}$ and ${0_{n \times m}}$ stands for unit matrix and null matrix respectively with compatible dimensions. 
Given two sets ${S_1}$ and ${S_1}$ in a Euclidean space, the operation $ \oplus $ stands for the Minkowski sum of ${S_1}$ and ${S_1}$. 
The operation $\otimes$ stands for Kronecker product.
The kernel is denoted by $\ker (A)$.
$R(A)$ stand for the row space of matrix $A$.
%
%
$A^\top$ stands for the transpose of $A$.
The interval hull of a bounded set $s = ({s^{(1)}}, \ldots {s^{(n)}}) \in \mathcal{S}$ is $\overline {{\text{IH}}} (\mathcal{S}) = \prod\nolimits_{i = 1}^n {[{{\underset{\raise0.3em\hbox{$\smash{\scriptscriptstyle-}$}}{s} }^{(i)}},{{\bar s}^{(i)}}]}$, where ${{\underset{\raise0.3em\hbox{$\smash{\scriptscriptstyle-}$}}{s} }^{(i)}} = {\inf _{{s^{(i)}}}}\{ s \in \mathcal{S}\}$ and ${{\bar s}^{(i)}} = {\sup _{{s^{(i)}}}}\{ s \in \mathcal{S}\}$.
A set ${\mathcal{S}_k}$ is uniformly bounded (w.r.t. $k \in \mathcal{K}$) if there exists a $\bar d > 0$ such that $D({\mathcal{S}_k}) \leqslant \bar d$ for all $k \in \mathcal{K}$.
\section{System Model and Problem Description}\label{sec_pb}
Consider a class of linear discrete-time systems at time $k \in {\mathbb{N}_0}$ modeled by uncertain variables as follows:
\begin{align}
{{\mathbf{x}}_{k + 1}} &= \Phi {{\mathbf{x}}_k} + \Gamma {u_k} + G{{\mathbf{d}}_k},\label{eq_ori_sysmodel}\\
{{\mathbf{y}}_k} &= \Xi {{\mathbf{x}}_k},\label{eq_ori_outmodel}
\end{align}
where \eqref{eq_ori_sysmodel} and \eqref{eq_ori_outmodel} are called the state and measurement equations, respectively, with $\Phi  \in {\mathbb{R}^{n \times n}}$, $\Gamma  \in {\mathbb{R}^{n \times p}}$, $G \in {\mathbb{R}^{n \times q}}$ and $\Xi  \in {\mathbb{R}^{l \times n}}$.
The system state is $\mathbf{x}_k$ (with its realization $x_k \in \llbracket\mathbf{x}_k\rrbracket \subseteq \mathbb{R}^n$); 
the disturbance is denoted by $\mathbf{d}_k$ (whose realization is $d_k \in \llbracket\mathbf{d}_k\rrbracket \subseteq \mathbb{R}^q$);
${u_k} \in {\mathbb{R}^p}$ is the control input; 
$\mathbf{y_k}$ is the measurement (with its realization ${y_k}\in\llbracket\mathbf{y_k}\rrbracket \subseteq {\mathbb{R}^l}$). 

In \eqref{eq_ori_sysmodel}, the disturbance is generated by the following exogenous system \cite{Su2018}:
\begin{equation}\label{eq_dismodel}
{{\mathbf{d}}_{k + 1}} = {{\mathbf{d}}_k} + \Delta {{\mathbf{d}}_k},
\end{equation}
where $\Delta {{\mathbf{d}}_k}$ (with its realization $\Delta d_k \in\llbracket \Delta{{\mathbf{d}}_k}\rrbracket\subseteq {\mathbb{R}^q}$) satisfies $D(\llbracket \Delta{{\mathbf{d}}_k}\rrbracket)\leqslant {D_\Delta }$, with $D_\Delta\in [0,\infty)$.
The initial value of the disturbance is arbitrary, i.e., $\llbracket \mathbf{d}_0\rrbracket = \mathbb{R}^q$.
Besides, we consider the following commonly used assumptions.
\begin{assumption}[Unrelatedness\!\cite{cong2022stability}]\label{asp_unrelated}
$\forall k \in {\mathbb{N}_0}$, ${{{\mathbf{x}}_0}}$, $\Delta {{\mathbf{d}}_{0:k}}$ are unrelated.
\end{assumption}
\begin{assumption}[\!\!\cite{Kim2013}]\label{asp_G}
Unknown input matrix $G$ has full column-rank, i.e., ${\text{Rank}}(G) = q$.
\end{assumption}
For the system described by~\eqref{eq_ori_sysmodel} and~\eqref{eq_ori_outmodel},
%
the disturbance observer (DO) is to estimate the disturbance $d_k$ based on all available control inputs and measurements up to $k$, i.e., $u_{0:k-1}, y_{0:k}$. 
Mathematically, the estimation of disturbance is $\hat{d}_k(u_{0:k-1}, y_{0:k})$ at $k \in \mathbb{N}_0$, and we write it as $\hat{d}_k(u_{0:k-1}, y_{0:k}) = \hat{d}_k$ for conciseness. 
A primary goal of DO is to make sure the estimation error (of the disturbance) $\hat{d}_k - d_k$ does not go unbounded as time elapses \cite{Kim2013,Su2018}. 
To support the theoretical results in the rest of this paper, the boundedness of the estimation error is rigorously defined in an asymptotic manner as follows.
\begin{definition}[Boundedness Of Disturbance Observer]\label{def_wellpest}	
A DO is bounded if the estimation error ${{{\hat d}_k} - {d_k}}$ satisfies
\begin{equation}\label{eq_boundedness_defination}
    \mathop {\overline {\lim } }\limits_{k \to \infty } \mathop {\sup }\limits_{{{d}_k} \in \llbracket {\mathbf{d}}_k \rrbracket^+}\left\| {{{\hat d}_k} - {d_k}} \right\| < \infty,
\end{equation}
where $\llbracket {\mathbf{d}}_k \rrbracket^+ := \llbracket {\mathbf{d}}_k|u_{0:k-1}, y_{0:k}\rrbracket$.
\end{definition}
With \defref{def_wellpest}, in this work, we focus on (i) analyzing when a bounded DO exists (see \probref{p1}) and (ii) proposing a bounded DO under the existence condition given by~(i) (see \probref{p2}).
\begin{problem}[Existence Condition]\label{p1}
Given the system described by \eqref{eq_ori_sysmodel} and \eqref{eq_ori_outmodel}, what is the necessary and sufficient condition for the existence of bounded DOs?
\end{problem}
%
%
\begin{problem}[Complete Algorithm]\label{p2}
How to design an efficient DO framework such that it is bounded if and only if the existence condition in \probref{p1} holds?
\end{problem}
To solve \probref{p1}, the necessary and sufficient existence condition of the bounded DOs is determined explicitly in \secref{sec_existence}, and it is compared with related work in \secref{sec_relationshipexisting}. 
To tackle \probref{p2}, in \secref{sec_algorithm}, a complete algorithm is proposed based on SMF, with proved worst-case optimality.
Finally, a simulation case subjected to the unbounded disturbance in \secref{sec_num} is performed to validate the theoretical results in \secref{sec_existence_analysis} and \secref{sec_algorithm}.
\section{Analysis of the Existence Condition}\label{sec_existence_analysis}

In this section, firstly, we establish a necessary and sufficient existence condition of a bounded DO in \secref{sec_existence} (see \thmref{thm_onlycondi}), which solves \probref{p1}.
Secondly, in \secref{sec_relationshipexisting} comparisons between our provided condition and the state-of-the-art results are made.

\subsection{Existence Condition Of Bounded Disturbance Observer}\label{sec_existence}

To begin with, we provide \lemref{lm_boundness tranfer} to show that the existence of bounded DOs is equivalent to the boundedness of $\llbracket\mathbf{d}_k\rrbracket^+$.
\begin{lemma}\label{lm_boundness tranfer}	
A bounded DO exists if and only if $\llbracket{{{\mathbf{d}}_k}}\rrbracket^+$ is uniformly bounded (w.r.t. $k \in {\mathbb{N}_0}$).
\end{lemma}
\begin{IEEEproof}
\emph{Sufficiency:} If $\llbracket{{{\mathbf{d}}_k}}\rrbracket^+$ is uniformly bounded, then $\exists$ ${{\hat d}_k} \in\llbracket{{{\mathbf{d}}_k}}\rrbracket^+$ such that
\begin{equation}\label{key}
    \mathop {\overline {\lim } }\limits_{k \to \infty } \mathop {\sup }\limits_{{{d}_k} \in \llbracket{{{\mathbf{d}}_k}}\rrbracket^+} \left\| {{{\hat d}_k} - {d_k}} \right\| \leqslant  D(\llbracket{{{\mathbf{d}}_k}}\rrbracket^+) < \infty .
\end{equation}

\emph{Necessity:} By the contrapositive, if $\llbracket{{{\mathbf{d}}_k}}\rrbracket^+$ is unbounded, then $\varlimsup_{k \to \infty} D(\llbracket{{{\mathbf{d}}_k}}\rrbracket^+) = \infty$ and
\begin{equation}\label{key}
    \begin{split}
        &\mathop {\sup }\limits_{d_k^1,d_k^2 \in \llbracket{{{\mathbf{d}}_k}}\rrbracket^+} \left\| {d_k^1 - d_k^2} \right\| \\&= \mathop {\sup }\limits_{d_k^1,d_k^2 \in \llbracket{{{\mathbf{d}}_k}}\rrbracket^+} \left\| {d_k^1 - d_k^2 + {{\hat d}_k} - {{\hat d}_k}} \right\|\\
        &\mathop  \leqslant \limits \mathop {\sup }\limits_{d_k^1 \in \llbracket{{{\mathbf{d}}_k}}\rrbracket^+} \left\| {d_k^1 - {{\hat d}_k}} \right\| + \mathop {\sup }\limits_{d_k^2 \in \llbracket{{{\mathbf{d}}_k}}\rrbracket^+} \left\| {d_k^2 - {{\hat d}_k}} \right\|\\
        &\leqslant 2\mathop {\sup }\limits_{{d_k} \in \llbracket{{{\mathbf{d}}_k}}\rrbracket^+} \left\| {{d_k} - {{\hat d}_k}} \right\|.
    \end{split}
\end{equation}
Therefore, $\forall{{\hat d}_k}$ one has that
\begin{equation}\label{key}
    \mathop {\overline {\lim } }\limits_{k \to \infty } \mathop {\sup }\limits_{{d_k} \in \llbracket{{{\mathbf{d}}_k}}\rrbracket^+} \left\| {{{\hat d}_k} - {d_k}} \right\| = \infty,
\end{equation}
which means bounded DOs do not exist.	
\end{IEEEproof}
According to \lemref{lm_boundness tranfer}, there are two issues (\textbf{i}. determining the actual disturbance range $\llbracket{{{\mathbf{d}}_k}}\rrbracket^+$; \textbf{ii}. proving the boundedness of $\llbracket{{{\mathbf{d}}_k}}\rrbracket^+$) for the establishment of the existence condition of bounded DOs (see the proof of \thmref{thm_onlycondi} in \apxref{apx_onlycondi}).
To better support the establishment of \thmref{thm_onlycondi}, we introduce the augmented form of \eqref{eq_ori_sysmodel}-\eqref{eq_dismodel}:
\begin{align}
{{\mathbf{z}}_{k + 1}} &= A {{\mathbf{z}}_k} + \bar \Gamma {u_k} + \Delta {\mathbf{\bar d}}_k,\label{eq_au_sysmodel}\\
{{\mathbf{y}}_k} &= C {{\mathbf{z}}_k},\label{eq_au_outmodel}
\end{align}
where ${{\mathbf{z}}_k} = {[{\mathbf{x}}_k^\top,{\mathbf{d}}_k^\top]}^\top$ combines system states with disturbance. 
Besides, the system parameters and $\Delta {\mathbf{\bar d}}_k$ are as follows:
\begin{equation}\label{eq_coeffi}
\begin{gathered}
    A =  {\begin{bmatrix}
            \Phi &G \\ 
            {{0_{q \times n}}}&{{I_{q\times q}}} 
    \end{bmatrix}},\quad {\bar \Gamma } = {\begin{bmatrix}
            \Gamma  \\ 
            {{0_{q \times p}}} 
    \end{bmatrix}} , \hfill \\
    C = \begin{bmatrix}
        \Xi &{{0_{l \times q}}} 
    \end{bmatrix},\quad \Delta {\mathbf{\bar d}}_k ={\begin{bmatrix}
  {{0_{n \times q}}} \\ 
  {{I_{q \times q}}} 
\end{bmatrix}}\Delta {{\mathbf{d}}_k}. \hfill \\ 
\end{gathered}
\end{equation}
Then, applying observability decomposition to \eqref{eq_au_sysmodel}-\eqref{eq_au_outmodel}, we have
\begin{equation}\label{eq_augaug_sys}
{{{\mathbf{\tilde z}}}_{k + 1}} = \tilde A{{{\mathbf{\tilde z}}}_k} + \tilde \Gamma {u_k} + \tilde B\Delta {{\mathbf{d}}_k},
\end{equation}
\begin{equation}\label{eq_augaug_syso}
{{\mathbf{y}}_k} = \tilde C{{{\mathbf{\tilde z}}}_k},
\end{equation} 
where 
\begin{equation}
    \begin{gathered}
  \tilde A =TA{T^{ - 1}} = {\begin{bmatrix}
  {{{\tilde A}_o}}&0 \\ 
  {{{\tilde A}_{21}}}&{{{\tilde A}_{\bar o}}} 
\end{bmatrix}},\tilde B =  {\begin{bmatrix}
  {{{\tilde B}_o}} \\ 
  {{{\tilde B}_{\bar o}}} 
\end{bmatrix}}  = T{\begin{bmatrix}
  {{0_{n \times q}}} \\ 
  {{I_{q \times q}}} 
\end{bmatrix}} \hfill \\
  \tilde \Gamma  = {\begin{bmatrix}
  {\tilde \Gamma _o^ \top }&{\tilde \Gamma _{\bar o}^ \top } 
\end{bmatrix}}^ \top= T\bar \Gamma ,\tilde C =  {\begin{bmatrix}
  {{{\tilde C}_o}}&0 
\end{bmatrix}}=C{T^{ - 1}}; \hfill \\ 
\end{gathered}
\end{equation}
${{{\mathbf{\tilde z}}}_k} = {[\begin{array}{*{20}{c}}
  {{{({\mathbf{\tilde z}}_k^o)}^ \top }}&{{{({\mathbf{\tilde z}}_k^{\bar o})}^ \top }} 
\end{array}]^ \top } = T{{\mathbf{z}}_k}$, and $T = {\begin{bmatrix}
    {{T_o}^\top}&{{T_{\bar o}}^\top} 
\end{bmatrix}^\top}$ is the transformation matrix, with ${T_o} \in {\mathbb{R}^{N_o \times (n + q)}}$, and ${T_{\bar o}} \in {\mathbb{R}^{N_{\bar o} \times (n + q)}}$;
Now, we give \thmref{thm_onlycondi} which is the necessary and sufficient existence condition of a bounded DO based on \eqref{eq_au_sysmodel} and \eqref{eq_au_outmodel}.
\begin{theorem}[Existence condition of bounded DO]\label{thm_onlycondi}
There exists a bounded DO if and only if the system satisfies:
\begin{equation}\label{eq_thm2}
    {\mathrm{rank}}( {\begin{bmatrix}
            O&{\bar IOG} 
    \end{bmatrix}}) = {\mathrm{rank}}( O ) + q,
\end{equation} 
where 
\begin{equation}\label{eq_thm2_para}
    O = {\begin{bmatrix}
            \Xi  \\ 
            {\Xi \Phi } \\ 
            \vdots  \\ 
            {\Xi {\Phi ^{n - 1}}} 
    \end{bmatrix}} ,\quad \bar I =  {\begin{bmatrix}
            0&0& \cdots &0 \\ 
            1&0& \cdots &0 \\ 
            \vdots & \ddots & \ddots & \vdots  \\ 
            1& \cdots &1&0 
    \end{bmatrix}} \otimes {I_{l \times l}}.
\end{equation}
\end{theorem}
\begin{IEEEproof}
See~\apxref{apx_onlycondi}.
\end{IEEEproof}
\begin{remark}
From \thmref{thm_onlycondi}, we can clearly see that a bounded DO exists if and only if $N_{ o}= n_o+q$, where $n_o$ is the dimension of the observable subsystem of \eqref{eq_ori_sysmodel}.
\end{remark}
%
%
Moreover, the following corollary can be derived from \thmref{thm_onlycondi}.
\begin{corollary}\label{cor_thm}
Bounded DOs do not exist for the system described by \eqref{eq_ori_sysmodel} and \eqref{eq_ori_outmodel} if $n_o<q$.
\end{corollary}
\begin{IEEEproof}
By \thmref{thm_onlycondi}, a bounded DO exists if and only if \eqref{eq_thm2} is satisfied. It implies that the system should satisfy the necessary condition
\begin{equation}\label{eq_co_nec}
    {\mathrm{rank}}(\bar IOG)\geqslant q.
\end{equation}
Besides, since 
\begin{equation}\label{key}
    {\mathrm{rank}}(\bar IOG) \leqslant \min {({\mathrm{rank}}(\bar I),{\mathrm{rank}}(O),{\mathrm{rank}}(G))}=n_o,
\end{equation} we must have $n_o \geqslant q$ to satisfy \eqref{eq_co_nec}. 
Therefore, for any system with $n_o<q$, bounded DOs do not exist.
\end{IEEEproof}
\begin{remark}
From \corref{cor_thm}, we can see that if bounded DOs exist, the dimension of the observable subsystem of \eqref{eq_ori_sysmodel} should be larger than the dimension of disturbance. Thus, given a control system, its observable states determine the maximum number of disturbances that can be estimated with bounded error.
\end{remark}
The existence condition provided by \thmref{thm_onlycondi} is explicit and can be easily verified based on the parameters of the original system. 
Thus, \probref{p1} has been solved. 
In the following subsection, we compare our necessary and sufficient condition (in \thmref{thm_onlycondi}) with the existing results.
\subsection{Comparison To Existing Results}\label{sec_relationshipexisting}
In this subsection, we compare the existence condition in \thmref{thm_onlycondi} and the results in~\cite{chang2006applying},~\cite{Kim2013} and~\cite{Su2018}. 

In \cite{chang2006applying}, the authors provide an existence condition for simultaneously estimating the system state and the disturbance, i.e., the pair $(A, C)$ is observable. This condition turns out to be a sufficient one for the existence of bounded DOs. More specifically, since the pair ($A$,$C$) is observable, we have $\mathrm{rank}([O~\bar{I}OG]) = n + q$, which implies \eqref{eq_thm2} is satisfied, but the converse is not necessarily true.

References~\cite{Kim2013} and~\cite{Su2018} provide different existence conditions for their designed DOs, respectively. 
The results in~\cite{Su2018} show that its existence condition is less conservative than that in~\cite{Kim2013}. 
This is because the DO proposed in~\cite{Su2018}, called SFO-based DO, is a generalized version of that in~\cite{Kim2013}. 
Therefore, we only discuss the relationship between our results and the results in \cite{Su2018}. 
The existence condition of SFO-based DOs in \cite{Su2018} is related to a state function (i.e. $z=Lx$) that is designed by a search method (i.e. the selection of $L$ starts with a low order in the complement space of $C$ and then increases the order until the existence condition are satisfied), while the existence of $L$ has not been proved. 
This means the existence condition in \cite{Su2018} is not necessary and sufficient for the existence of bounded DOs. 
Besides, even though $L$ exists, the condition provided in~\cite{Su2018} is not explicit and hard to verify;
this is because the existence condition depends not only on the system parameters but also on the specific form of $L$ to be designed. 
In contrast, our established existence condition is a necessary and sufficient one, which is only related to the system parameters, and is easy to verify.

\section{Bounded and Optimal Disturbance Observer: A Set-Membership Filter-Based Method}\label{sec_algorithm}

In this section, we focus on \probref{p2}.
More specifically, a reduced-order SMF-based DO (SMFDO) is proposed in \secref{subsecalgorithm}, which is proved to be bounded in \secref{subsecalgoboundedeneess}.
Furthermore, we prove the optimality of the proposed SMFDO in \secref{secsuboptimal}.
\subsection{Set-Membership Filter-Based Disturbance Observer}\label{subsecalgorithm}

Our proposed SMFDO is based on the Constrained Zonotope (CZ).
To start with, we provide the definition of CZ as follows.
\begin{definition}[\!\!\cite{cong2022stability}]\label{def_constrainedzonotopic}
A set $\mathcal{Z} \subseteq {\mathbb{R}^n}$ is a (extended) constrained zonotope if there exists a quintuple $(\hat G,\hat c,\hat A,\hat b, \hat h) \in {\mathbb{R}^{n \times {n_g}}} \times {\mathbb{R}^n} \times {\mathbb{R}^{{n_c} \times {n_g}}} \times {\mathbb{R}^{{n_c}}}\times [0,\infty]^{n_g}$ such that $\mathcal{Z}$ is expressed by
\begin{equation}\label{eq_constrainzono}
    \left\{ {\hat G\xi  + \hat c:\hat A\xi  = \hat b,\xi  \in \prod\limits_{j = 1}^{{n_g}} {[ - {{\hat h}^{(j)}},{{\hat h}^{(j)}}]} } \right\} = :Z(\hat G,\hat c,\hat A,\hat b,\hat h),
\end{equation}
where ${\hat h}^{(j)}$ is the $j$-th component of $\hat h$.
\end{definition}

\begin{remark}\label{rek:CZ}
The reason why we use CZ is two-fold: it is suitable for most of the disturbance models, including the classical one\footnote{According to the previous work \cite{Kim2013} and \cite{Su2018}, $\Delta {{{d}}_k}$ can be modeled by the following componentwise range $\Delta {\mathbf{d}}_{k + 1}^i = \left| {{{d}}_{k + 1}^i - {{d}}_k^i} \right| \leqslant {\bar d}$, 
    with ${\bar d}$ a small positive value.
} in~\cite{Kim2013} and~\cite{Su2018};
for any bounded disturbance models, we can always find a CZ as an outer bound of $\llbracket {\Delta {{\mathbf{d}}_k}} 
\rrbracket$.
\end{remark}
Then, we provide the reduced-order SMFDO in \algref{alg:thm}, and the line-by-line explanation as follows.
\begin{algorithm}
\begin{footnotesize}
    \caption{SMF-based Disturbance Observer}\label{alg:thm}
    \begin{algorithmic}[1]
        \STATE  \textbf{Initialization:} 
        Transformation matrix $T$;           
        system parameters ${{{\tilde A}_o}}$, ${{{\tilde B}_o}}$, ${{{\tilde \Gamma}_o}}$ and ${{{\tilde C}_o}}$; 
        initial disturbance range $\llbracket {{{\mathbf{d}}_0}} 
        \rrbracket = {\mathbb{R}^q}$;
        initial state range $\llbracket {{{\mathbf{x}}_0}} 
        \rrbracket\subseteq  {\mathbb{R}^n}$;
        filtering interval $\delta>\mu-1$.
        \STATE $\mathcal{Z}_{0}^- =Z( \hat G_0^ - , \hat c_0^ - ,\hat A_0^ - , \hat b_0^ -,{\hat h_{0}^ -} ) \supseteq {T_o}(\llbracket {{{\mathbf{x}}_0}} 
        \rrbracket \times \llbracket {{{\mathbf{d}}_0}} 
        \rrbracket)$;
        \STATE Solve the linear map $P$ from $PT_o=\begin{bmatrix}
            {{0_{q \times n}}}&{{I_{q \times q}}} 
        \end{bmatrix}$;
        \LOOP
        \IF {$k < \delta$}
        \IF {$k > 0$}
        \STATE $\mathcal{Z}_{k}^{\Delta}=Z( \hat G_k^ \Delta , \hat c_k^ \Delta ,\hat A_k^ \Delta , \hat b_k^ \Delta,{\hat h_{k}^ \Delta} )\supseteq\llbracket {\Delta {{\mathbf{d}}_k}} \rrbracket$;
        \STATE $\mathcal{Z}_{k}^-= Z( \hat G_k^ - , \hat c_k^ - ,\hat A_k^ - , \hat b_k^ -,{\hat h_{k}^ -} )$ with
        \begin{equation}\label{eq_CZ_predict}
            \begin{gathered}
                \hat G_k^ -  =  {\begin{bmatrix}
                        {{{\tilde A}_o}{{\hat G}_{k - 1}}}&{{{\tilde B}_o}\hat G_{k - 1}^\Delta } 
                \end{bmatrix}},\hfill \\c_k^ -  = {{\tilde A}_o}{{\hat c}_{k - 1}} + \hat c_{k - 1}^\Delta  + {{\tilde \Gamma}_o}{u_{k-1}}, \hfill \\
                \hat A_k^ -  =  {\begin{bmatrix}
                        {{{\hat A}_{k - 1}}}&0 \\ 
                        0&{\hat A_{k - 1}^\Delta } 
                \end{bmatrix}},\hat b_k^ -  = {\begin{bmatrix}
                        {{{\hat b}_{k - 1}}} \\ 
                        {\hat b_{k - 1}^\Delta } 
                \end{bmatrix}} ,\hat h_k^ -  = {\begin{bmatrix}
                        {{{\hat h}_{k - 1}}} \\ 
                        {\hat h_{k - 1}^\Delta } 
                \end{bmatrix}}. \hfill \\ 
            \end{gathered} 
        \end{equation}
        \ENDIF
        \STATE $\mathcal{Z}_{k}=Z({\hat G_{k}},{\hat c_{k}},{\hat A_{k}},{\hat b_{k}},{\hat b_{k}},{\hat h_{k}})$, with
        \begin{equation}\label{eq_CZ_update}
            \begin{gathered}
                {{\hat G}_k} = \hat G_k^ - ,\quad {{\hat c}_k} = \hat c_k^ - ,{{\hat h}_k} = \hat h_k^ -  \hfill \\
                {{\hat A}_k} =  {\begin{bmatrix}
                        {\hat A_k^ - } \\ 
                        {{{\tilde C}_o}\hat G_k^ - } 
                \end{bmatrix}} ,{b_k} =  {\begin{bmatrix}
                        {\hat b_k^ - } \\ 
                        {{y_k} - {{\tilde C}_o}\hat c_k^ - } 
                \end{bmatrix}};\hfill \\ 
            \end{gathered} 
        \end{equation}
        \ELSE 
        \STATE  $\mathcal{Z}_{k}=Z({\hat G_{k}},{\hat c_{k}},{\hat A_{k}},{\hat b_{k}},{\hat b_{k}},{\hat h_{k}})$, with the CZ parameter defined by \eqref{eq_CZPARA};
        \ENDIF
        \STATE$\mathcal{Z}_{k}^d = P\mathcal{Z}_{k}$;
        \STATE  ${{\hat d}_k} = {\text{center}}(\overline {{\text{IH}}}( \mathcal{Z}_{k}^d ) )$;
        \ENDLOOP
    \end{algorithmic}
\end{footnotesize}
\end{algorithm}

Line 1 initializes the transformation matrix, the system parameters, the initial state range, and the initial disturbance range. 
The choice of the filtering interval $\delta$ is supposed to satisfy $\delta \geqslant{\mu _o} - 1$, where ${\mu _o}$ is the observability index of the pair $(A, C)$. 
Larger $\delta$ brings more accurate estimation while increasing the computational complexity.

Line 2 gives the estimate $\mathcal{Z}_{0}^-$, which is the observable part of the Cartesian product of
the initial system state range and disturbance state range.
Line~3 obtains the linear map $P$ by solving $PT_o=\begin{bmatrix}
{{0_{q \times n}}}&{{I_{q \times q}}} 
\end{bmatrix}$.	
Lines 5-11 give the estimate $\mathcal{Z}_{k}$ when $k < \delta$. 
Specifically, line 7 chooses a CZ set $\mathcal{Z}_k^\Delta$ that can outer bound $\llbracket {\Delta {{\mathbf{d}}_k}} 
\rrbracket$.
Line 8 gives prior estimate $\mathcal{Z}_{k}^-$, where the quintuple $({\hat G_{k}},{\hat c_{k}},{\hat A_{k}},{\hat b_{k}},{\hat h_{k}})$ is defined by \eqref{eq_CZ_predict}.
Line 10 provides the posterior estimate $\mathcal{Z}_{k}$, with the CZ parameter determined by \eqref{eq_CZ_update}.
Different from Lines 5-11, Lines 12-13 provide the posterior estimate $\mathcal{Z}_{k}$ when $k \geqslant \delta$, where the quintuple of $\mathcal{Z}_{k}$ is defined by \eqref{eq_CZPARA}, with \[\hat b_{(l)}^y = {y_{k - \delta+1  + l}} - {{{\tilde C}_o}}[{{\tilde A}_o}^{l+1} {{\hat c}_{k - \delta }} + \sum\limits_{i = 0}^{l} {{{\tilde A}_o}^i(\hat c_{k - \delta +i}^\Delta  + {{\tilde \Gamma}_o}{u_{k - \delta +i}})}]\] for $0\leqslant l\leqslant\delta$. 
Specifically, the quintuple in \eqref{eq_CZPARA} is derived by recursively using \eqref{eq_CZ_predict} and \eqref{eq_CZ_update} $\delta$ times, and we let the interval hull of $\mathcal{Z}_{k-\delta}$ be the initial posterior range of the recursive process.	
Line~14 obtains the estimate $\mathcal{Z}_{k}^d$ by applying linear map $P$ on $\mathcal{Z}_{k}$.
Line~15 chooses the center of the interval hull of $\mathcal{Z}_{k}^d$ as the estimate of disturbance.
\begin{remark}
From the \algref{alg:thm}, we know that the estimation of disturbance is only related to the observable part of the system \eqref{eq_au_sysmodel} and \eqref{eq_au_outmodel}, which is illustrated in proof of \thmref{thm_onlycondi} [see \eqref{eq_thm2pf_transform1} in \apxref{apx_onlycondi}]. This implies that SMFDO is a reduced-order DO compared to other full-order DOs.
\end{remark}
\subsection{Boundedness of SMFDO}\label{subsecalgoboundedeneess}
In this subsection, we focus on the boundedness of \algref{alg:thm}, which is illustrated by the following theorem.
\begin{theorem}\label{thm_OIT_bounded}
The estimate ${\hat{d}}_k$ derived from \algref{alg:thm} satisfies~\eqref{eq_boundedness_defination}, i.e., the proposed DO is bounded,
if the system satisfies the existence condition in \thmref{thm_onlycondi}.
\end{theorem}
\begin{IEEEproof}
See~\apxref{apx_thm2}.
\end{IEEEproof}
From \thmref{thm_OIT_bounded}, \algref{alg:thm} has a bounded disturbance estimation error for $k \in {\mathbb{N}_0}$, which guarantees the boundedness of \algref{alg:thm}.
\subsection{Optimality of SMFDO}\label{secsuboptimal}
In this subsection, we prove that the proposed bounded DO framework is worst-case optimal. 
Firstly, the optimality is defined as follows.

\begin{definition}\label{definop}
An estimated disturbance $\hat{d}_k \in \mathbb{R}^q$ is optimal for system \eqref{eq_ori_sysmodel} and \eqref{eq_ori_outmodel}, if for any $\hat{d}_k' \in \mathbb{R}^q$, the following holds
\begin{equation}\label{eqn:Optimal Estimate}
    \sup_{d_k \in \llbracket\mathbf{d}_k\rrbracket^+} \|\hat{d}_k - d_k\| \leq \sup_{d_k \in \llbracket\mathbf{d}_k\rrbracket^+} \|\hat{d}_k' - d_k\|.
\end{equation}
\end{definition}

\defref{definop} gives the optimality of disturbance estimation in a worst-case manner:
since the real disturbance can be any point in $\llbracket{{{\mathbf{d}}_k}} \rrbracket^+$, the selected optimal estimate ${\hat d_k}$ should minimize the worst-case (maximum) estimation error.
The following theorem tells when \algref{alg:thm} is optimal.



%
\begin{theorem}\label{thm:Worst-Case Optimality}
The disturbance estimation $\hat d_k$ derived from \algref{alg:thm} is optimal if $k<\delta$ and
\begin{equation}\label{eq_optimal_re1}
    \mathcal{Z}_{0}^- = {T_o}(\llbracket {{{\mathbf{x}}_0}} 
        \rrbracket \times \llbracket {{{\mathbf{d}}_0}} 
        \rrbracket),\quad \mathcal{Z}_{k}^{\Delta}= \llbracket {\Delta {{\mathbf{d}}_k}} 
\rrbracket.
\end{equation}
\end{theorem}
\begin{IEEEproof}
See~\apxref{apx:Proof of thm:Worst-Case Optimality}.
\end{IEEEproof}

\thmref{thm:Worst-Case Optimality} implies that the uncertainties should be constrained zonotopes such that~\eqref{eq_optimal_re1} holds (see also \rekref{rek:CZ}).
In addition, $k$ should be less than $\delta$, which indicates: for small $k$, the optimality is achieved; for large $k$, the filtering interval $\delta$ should be large enough\footnote{It implies the computing power should be strong enough to guarantee the worst-case optimality. But it should be noted that even for small $\delta$, the estimation error of \algref{alg:thm} with such $\delta$ is very close to that of the worst-case optimal one [see \figref{sf2}].}.
Therefore, \algref{alg:thm} has the capability to achieve the worst-case optimality.

\begin{figure*}[ht] 
\begin{equation}\label{eq_CZPARA}
    \begin{gathered}
        \hat G_k  = \left[ {\begin{array}{*{20}{c}}
                {{{\tilde A}_o}^\delta }&{{{\tilde A}_o}^{\delta  - 1}{{\tilde B}_o}\hat G_{k - \delta }^\Delta }&{{{\tilde A}_o}^{\delta  - 2}{{\tilde B}_o}\hat G_{k - \delta  + 1}^\Delta }& \cdots &{{{\tilde B}_o}\hat G_{k - 1}^\Delta } 
        \end{array}} \right],\hat c_k  = {{\tilde A}_o}^\delta {{\hat c}_{k - \delta }} + \sum\limits_{i = 0}^{\delta  - 1} {{{\tilde A}_o}^i(\hat c_{k - i - 1}^\Delta  + {{\tilde \Gamma}_o}{u_{k - i - 1}})} , \hfill \\
        \hat A_k  = \left[ {\begin{array}{*{20}{c}}
                
                0&{\hat A_{k - \delta }^\Delta }&0& \cdots &0 \\ 
                {{{\tilde C}_o}{{\tilde A}_o}}&{{{\tilde C}_o}{{\tilde B}_o}\hat G_{k - \delta }^\Delta }&0& \cdots &0 \\ 
                0&0&{\hat A_{k - \delta  + 1}^\Delta }& \cdots &0 \\ 
                {{{\tilde C}_o}{{\tilde A}_o}^2}&{{{\tilde C}_o}{{\tilde A}_o}{{\tilde B}_o}\hat G_{k - \delta }^\Delta }&{{{\tilde C}_o}{{\tilde B}_o}\hat G_{k - \delta  + 1}^\Delta }& \cdots &0 \\ 
                \vdots & \vdots & \vdots &{}& \vdots  \\ 
                0&0&0& \cdots &{\hat A_{k - 1}^\Delta } \\ 
                {{{\tilde C}_o}{{\tilde A}_o}^\delta}&{{{\tilde C}_o}{{\tilde A}_o}^{\delta  - 1}{{\tilde B}_o}\hat G_{k - \delta }^\Delta }&{{{\tilde C}_o}{{\tilde A}_o}^{\delta  - 2}{{\tilde B}_o}\hat G_{k - \delta  + 1}^\Delta }& \cdots &{{{\tilde C}_o}{{\tilde B}_o}\hat G_{k - 1}^\Delta } 
        \end{array}} \right],\hat b_k  = \left[ {\begin{array}{*{20}{c}}
                
                {\hat b_{k - \delta }^\Delta } \\ 
                {\hat b_{(0)}^y} \\ 
                {\hat b_{k - \delta  + 1}^\Delta } \\ 
                {\hat b_{(1)}^y} \\ 
                \vdots  \\ 
                {\hat b_{(\delta )}^y} 
        \end{array}} \right],\hat h_k  = \left[ {\begin{array}{*{20}{c}}
                {{{\hat h}_{k - \delta }}} \\ 
                {\hat h_{k - \delta }^\Delta } \\ 
                \vdots  \\ 
                {{{\hat h}_k}} \\ 
                {\hat h_k^\Delta } 
        \end{array}} \right] \hfill \\				
    \end{gathered} 
\end{equation}
\end{figure*}
\section{Numerical Examples}\label{sec_num}
In this section, we consider a simulation case subjected to unbounded disturbance to verify the boundedness and optimality of SMFDO.
\begin{figure}[ht]	
\centering  
\subfigbottomskip=2pt 
\subfigcapskip=-5pt 
\subfigure[\label{sf1} ]{
    \includegraphics[width=0.9\linewidth]{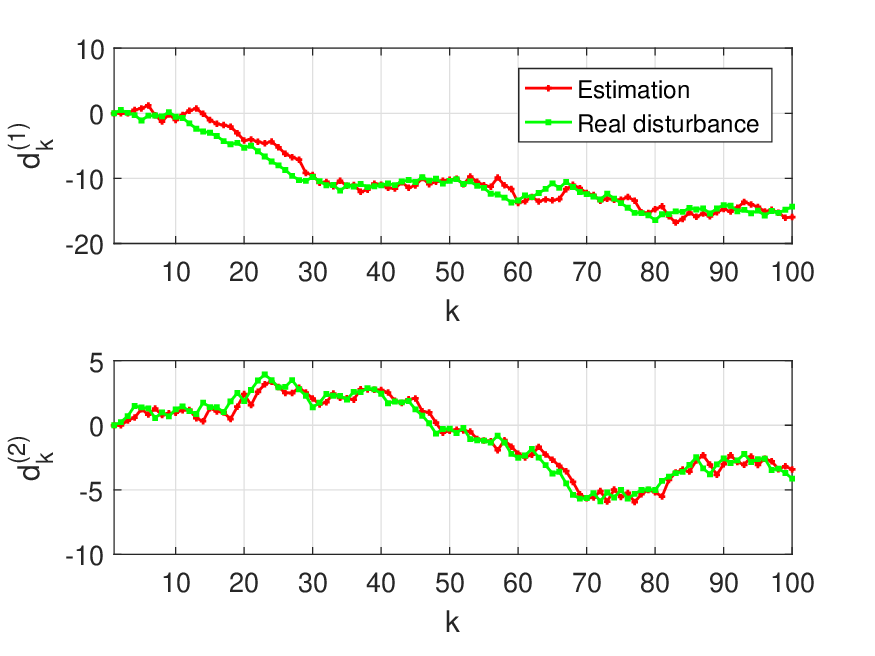}}   \hspace{-2.7mm}
\subfigure[\label{sf2}]{
    \includegraphics[width=0.92\linewidth]{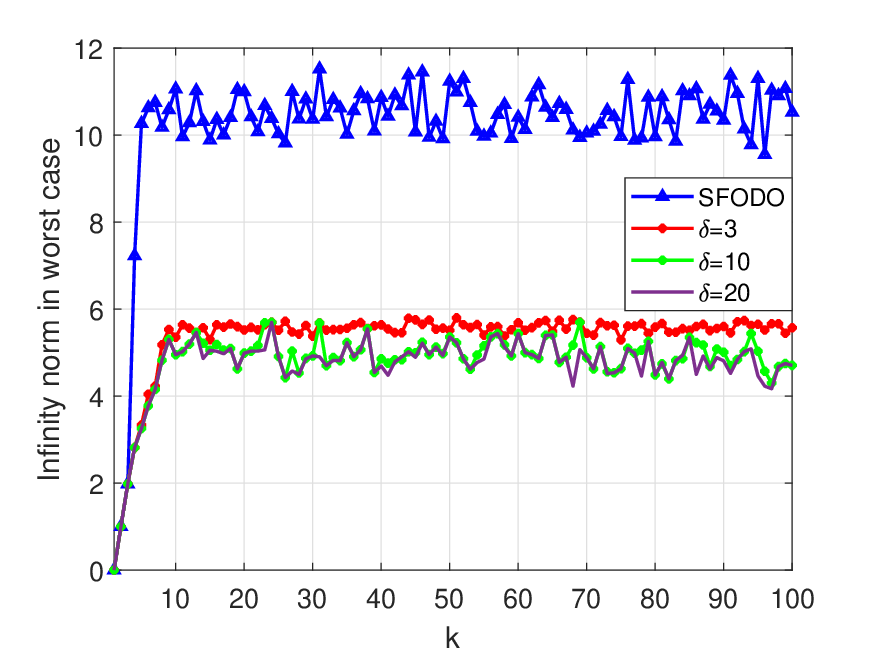}}  \hspace{-2.7mm}
\caption{Performance of the SMFDO: (a) the estimated disturbance given by SMFDO and the real disturbance over $k \in [0,100]$;
%
(b) comparison (w.r.t. the infinite norm of estimation error, i.e., $\|\hat{d}_k - d_k\|_{\infty}$) between SFODO and SMFDO under different $\delta$ using 5000 Monte Carlo simulation runs.}
\end{figure}
To better compare with previous results, we take the system in~\cite{chang2006applying} as an example, which is also used in \cite{Kim2013} and \cite{Su2018}.
The system parameters are as follows.
\begin{equation}\label{eq_case1}
\begin{gathered}
    \begin{gathered}
        \Phi  = {\begin{bmatrix}
                {0.9630}&{{\text{0}}{\text{.0181}}}&{{\text{0}}{\text{.0187}}} \\ 
                {{\text{0}}{\text{.1808}}}&{{\text{0}}{\text{.8195}}}&{{\text{-0}}{\text{.0514}}} \\ 
                {{\text{ -0}}{\text{.1116}}}&{{\text{0}}{\text{.0344}}}&{{\text{0}}{\text{.9586}}} 
        \end{bmatrix}}, \quad \Gamma  = {0_{3 \times 1}},\\
        G = {\begin{bmatrix}
                {{\text{0}}{\text{.0996}}}&{{\text{0}}{\text{.0213}}} \\ 
                {{\text{0}}{\text{.0050}}}&{{\text{0}}{\text{.1277}}} \\ 
                {{\text{0}}{\text{.1510}}}&{{\text{0}}{\text{.0406}}} 
        \end{bmatrix}}, \quad \Xi  =  {\begin{bmatrix}
                1&0&{ - 1} \\ 
                { - 1}&1&1 
        \end{bmatrix}}. \hfill \\ 
    \end{gathered}
\end{gathered}
\end{equation}
The disturbance is generated by the following exosystem:
\begin{equation}\label{eq_case2_dis}
{{\mathbf{d}}_{k + 1}} = {{\mathbf{d}}_k} + \Delta {{\mathbf{d}}_k},
\end{equation}
where $\Delta {{\mathbf{d}}_k}$ modeled by the following componentwise range
\begin{equation}
\Delta d_k^i = \left| {d_{k + 1}^i - d_k^i} \right| \leqslant 1.
\end{equation}

We compare the performance of the present SMFDO with SFODO in \cite{Su2018}.
In terms of parameter design, for SMFDO, $\llbracket {\Delta{{\mathbf{d}}_k}} 
\rrbracket$ is outer bounded by $Z(\hat G^\Delta ,\hat c^\Delta ,\hat A^\Delta ,\hat b^\Delta,\hat h^\Delta )$, with
\begin{equation}\label{eq_case2_dis1}
{\hat G^\Delta } = {I_{2 \times 2}},{\hat c^\Delta } ={\begin{bmatrix}
        0 \\ 
        0 
\end{bmatrix}},{\hat A^\Delta } = [],{\hat b^\Delta} = [],\hat h^\Delta={\begin{bmatrix}
        1 \\ 
        1 
\end{bmatrix}}.
\end{equation}
For SFODO, we use the optimal parameters designed by \cite{chang2006applying} to realize the SFODO, where the equivalence was proved in \cite{Su2018}.

\figref{sf1} shows the estimation performance of SMFDO for each entry of ${{\mathbf{d}}_k}$; \figref{sf2} compares the estimation accuracy of SFODO (under different $\delta$) and SMFDO.
We can see that the proposed SMFDO has a bounded estimation error, and SMFDO improves the estimation accuracy over $40\%$, which verifies the optimality of the SMFDO.
Furthermore, the estimation accuracy of our proposed SMFDO does not experience a significant improvement from $\delta=3$ to $\delta=20$ (particularly from $\delta = 10$ to $\delta = 20$), which implies that even with small $\delta$, the SMFDO is very close to optimal.

\section{Conclusion}\label{sec:Conclusion}
In this paper, the boundedness of DOs has been studied for linear discrete-time systems. 
By analyzing through a set-membership viewpoint, a necessary and sufficient existence condition for bounded DO has been proposed.
With this existence condition, one can easily check whether bounded DOs exist.
%
Using the CZ representation, we have designed the SMFDO with proved completeness and worst-case optimality.
Monte Carlo simulations have been employed to validate the effectiveness of the theoretical results.
%
%
%
\appendices

\section{Proof of \thmref{thm_onlycondi}}\label{apx_onlycondi}
To start with, we provide the following lemmas to support our proof.
\begin{lemma}[Boundedness of estimation \cite{cong2022stability}]\label{lm_SMFstable}
    For the system described by \eqref{eq_au_sysmodel} and \eqref{eq_au_outmodel}, if the pair $(A,C)$ is observable, then	
    \begin{equation}
        \varlimsup_{k \to \infty} D(\llbracket{{{\mathbf{z}}_k}}\rrbracket^+) < \infty,
    \end{equation}
    where $\llbracket\mathbf{z}_k\rrbracket^+ := \llbracket\mathbf{z}_k|u_{0:k-1},y_{0:k}\rrbracket$.
\end{lemma}
\begin{lemma}\label{lm_eqsolotiuon}
    There exists a solution for the system of linear equations $AX = B$ if and only if:
    \begin{equation}\label{key}
        \mathrm{rank}(A) = \mathrm{rank}(\begin{bmatrix}
            A & B
        \end{bmatrix}).
    \end{equation}
\end{lemma}
\begin{lemma}\label{lm_unboundedcombine}
    For any ${{\mathbf{x}}_1}$ and ${{\mathbf{x}}_2}$ satisfying $D(\llbracket{{\mathbf{x}}_1}\rrbracket)=\infty$ and $D(\llbracket{{\mathbf{x}}_2}\rrbracket)<\infty$, the following property holds:
    \begin{equation}\label{key}
        D(\llbracket{{\mathbf{x}}_1}+{{\mathbf{x}}_2}\rrbracket)=\infty.
    \end{equation}
\end{lemma}
\begin{IEEEproof}
    Since $D(\llbracket{{\mathbf{x}}_1}\rrbracket)=\infty$, one has that $\forall \bar D > 0$, $\exists {\omega _1},{\omega _2} \in \Omega$ such that
    \begin{equation}\label{key}
        \left\| {{{\mathbf{x}}_1}({\omega _1}) - {{\mathbf{x}}_1}({\omega _2})} \right\| \geqslant \bar D.
    \end{equation}
    Let $D(\llbracket {{{\mathbf{x}}_2}} 
    \rrbracket) = {D_2}$, and then we have
    \begin{equation}\label{key}
        \begin{split}
            D(\llbracket {{{\mathbf{x}}_1} + {{\mathbf{x}}_2}} 
            \rrbracket) &\geqslant \left\| {{{\mathbf{x}}_1}({\omega _1}) + {{\mathbf{x}}_2}({\omega _2}) - {{\mathbf{x}}_1}({\omega _2}) - {{\mathbf{x}}_2}({\omega _1})} \right\|\\ 
            &\geqslant \left\| {{{\mathbf{x}}_1}({\omega _1}) - {{\mathbf{x}}_1}({\omega _2})} \right\| + \left\| {{{\mathbf{x}}_2}({\omega _1}) - {{\mathbf{x}}_2}({\omega _2})} \right\|\\
            &\geqslant \bar D - {D_2}.
        \end{split}
    \end{equation}
    As $\bar{D}$ is arbitrary, we get $D(\llbracket{{\mathbf{x}}_1}+{{\mathbf{x}}_2}\rrbracket)=\infty$.
\end{IEEEproof}
Recall that there are two issues for establishing the existence condition of bounded DOs. 
According to the structure of the augmented system described by \eqref{eq_au_sysmodel} and \eqref{eq_au_outmodel}, it is clear that the relationship between $\llbracket{{{\mathbf{d}}_k}}\rrbracket^+$ and $\llbracket{{{\mathbf{z}}_k}}\rrbracket^+$ can be expressed by (issue \textbf{i}) 
\begin{equation}\label{eq_transform_P}
    {\begin{bmatrix}
            {{0_{n \times q}}}&I_{q \times q} 
    \end{bmatrix}}\llbracket{{{\mathbf{z}}_k}}\rrbracket^+ = \llbracket{{{\mathbf{d}}_k}}\rrbracket^+.
\end{equation}
Meanwhile, \eqref{eq_transform_P} imples that the boundedness of $\llbracket{{\mathbf{d}}_k}\rrbracket^+$ is related to the boundedness of $\llbracket{\mathbf{z}}_k\rrbracket^+$ (issue \textbf{ii}).
The above analysis provides the following two steps to complete the proof, corresponding to the sufficiency and necessity of \thmref{thm_onlycondi} respectively.

\emph{Step~1 (Sufficiency):}
If \eqref{eq_thm2} is satisfied, then one has\footnote{According to \eqref{eq_thm2}, it is clear that ${\bar IOG}$ has full column rank, with $\mathrm{rank}(\bar IOG)=q$. Then, we can readily construct the matrix ${\begin{bmatrix}
            O&{\bar IOG} \\ 
            0&{{I_{q \times q}}} 
    \end{bmatrix}}$ such that \eqref{eq_thm2pf_transformconclu} is satisfied.}
\begin{equation}\label{eq_thm2pf_transformconclu}
    {\mathrm{rank}} {\begin{bmatrix}
            O&{\bar IOG} 
    \end{bmatrix}} = {\mathrm{rank}({\begin{bmatrix}
                O&{\bar IOG} \\ 
                0&{{I_{q \times q}}} 
        \end{bmatrix}})},
\end{equation}
which is equivalent to
\begin{equation}\label{eq_thm2pf_transform6}
    {\mathrm{rank}}(\left[ {\begin{array}{*{20}{c}}
            {{O^\top}}&0 \\ 
            {{{(\bar IOG)}^\top}}&{{I_{q \times q}}} 
    \end{array}} \right]) = {\mathrm{rank}}(\left[ {\begin{array}{*{20}{c}}
            {{O^\top}} \\ 
            {{{(\bar IOG)}^\top}} 
    \end{array}} \right]).
\end{equation}
According to \lemref{lm_eqsolotiuon} and \eqref{eq_thm2pf_transform6}, there exists a $P' \in {\mathbb{R}^{q \times (l \times n)}}$ such that
\begin{equation}\label{eq_thm2pf_transform5}
    {\begin{bmatrix}
            {{O^\top}} \\ 
            {{{(\bar IOG)}^\top}} 
    \end{bmatrix}}{{P'}^\top} = {\begin{bmatrix}
            0 \\ 
            {{I_{q \times q}}} 
    \end{bmatrix}}.
\end{equation}
Transposing both sides of \eqref{eq_thm2pf_transform5}, we have
\begin{equation}\label{eq_thm2pf_transform4}
    P'\begin{bmatrix}
        O&{\bar IOG} 
    \end{bmatrix} = \begin{bmatrix}
        0_{q \times n}&{{I_{q \times q}}} 
    \end{bmatrix}.
\end{equation}
According to \eqref{eq_thm2_para}, we establish the observability matrix of the augmented system \eqref{eq_au_sysmodel} as follows:
\begin{equation}\label{eq_aug_observematrix}
    {O_{aug}} = \left[ \begin{gathered}
        \begin{array}{*{20}{c}}
            \Xi &0 
        \end{array} \hfill \\
        \begin{array}{*{20}{c}}
            {\Xi \Phi }&{\Xi G} 
        \end{array} \hfill \\
        \begin{array}{*{20}{c}}
            {\Xi {\Phi ^2}}&{\Xi \Phi G + \Xi G} 
        \end{array} \hfill \\
        \begin{array}{*{20}{c}}
            {\Xi {\Phi ^3}}&{\Xi {\Phi ^2}G + \Xi \Phi G + \Xi G} 
        \end{array} \hfill \\
        \begin{array}{*{20}{c}}
            \vdots & \vdots  
        \end{array} \hfill \\
        \begin{array}{*{20}{c}}
            {\Xi {\Phi ^{n - 1}}}&{\sum\limits_{i = 0}^{n - 2} {\Xi {\Phi ^{n - 2}}G} } 
        \end{array} \hfill \\ 
    \end{gathered}  \right]= \begin{bmatrix}
        O&{\bar IOG} \end{bmatrix}.
\end{equation}
Let
\begin{equation}\label{eq_thm2pf_transform3}
    {T_o} = {T_c}{O_{aug}},
\end{equation}
where ${T_c} \in {\mathbb{R}^{o' \times (l \times n)}}$ stands for the linear map that chooses ${o'}$ linearly independent row vectors from ${O_{aug}}$.
Since $o' = \mathrm{rank}(O_{aug})$, there must exists a linear map ${T_c}^\prime  \in {\mathbb{R}^{(l \times n) \times o'}}$ such that
\begin{equation}\label{eq_thm2pf_transform7}
    {T_c}^\prime {T_o} = {O_{aug}}.
\end{equation}
Let $P = P'{T_c}^\prime$, then the following equation can be obtained:
\begin{equation}\label{eq_thm2pf_transform1}
    \begin{split}
        \llbracket{{{\mathbf{d}}_k}}\rrbracket^+&\mathop=\limits^{\eqref{eq_transform_P}} \begin{bmatrix}
            {{0_{q \times n}}}&{{I_{q \times q}}} 
        \end{bmatrix}\llbracket{{{\mathbf{z}}_k}}\rrbracket^+ \mathop=\limits^{\eqref{eq_thm2pf_transform4}} P' O_{aug}\llbracket{{{\mathbf{z}}_k}}\rrbracket^+
        \\&\mathop=\limits^{\eqref{eq_thm2pf_transform7}} P'{T_c}^\prime {T_o}\llbracket{{{\mathbf{z}}_k}}\rrbracket^+ = P\llbracket {{\mathbf{\tilde z}}_k^o} 
        \rrbracket^+.		
    \end{split}
\end{equation}
Then, according to \eqref{eq_thm2pf_transform1}, we have the following equation
\begin{equation*}
    \begin{split}
        &{\varlimsup_{k\to\infty}\sup _{{{\hat d}_k},{d_k} \in \llbracket{{{\mathbf{d}}_k}}\rrbracket^+}}\left\| {{{\hat d}_k} - {d_k}} \right\| = \varlimsup_{k\to\infty}{\sup _{\hat {\tilde z}_k^o,\tilde z_k^o \in \llbracket {\tilde{\mathbf{ z}}_k^o} 
                \rrbracket^+}}\left\| {P\hat {\tilde z}_k^o - P\tilde z_k^o} \right\|\\
        &\leqslant  \varlimsup_{k\to\infty}\left\| P \right\|{\sup _{\hat {\tilde z}_k^o,\tilde z_k^o \in \llbracket  {\tilde{\mathbf{ z}}_k^o} 
                \rrbracket^+ }}\left\| {\hat {\tilde z}_k^o - \tilde z_k^o} \right\|
        < \infty,
    \end{split}
\end{equation*}
which gives the uniform boundedness of $\llbracket{{{\mathbf{d}}_k}}\rrbracket^+$.

\emph{Step~2 (Necessity):}
By the contrapositive, if \eqref{eq_thm2} is not satisfied, from \lemref{lm_eqsolotiuon}, $P'$ in \eqref{eq_thm2pf_transform4} does not exist, which means there exists at least one row vetcor in ${\begin{bmatrix}
        0 & 
        {{I_{q \times q}}} 
\end{bmatrix}}$ that can be expressed by the complement space of $R(O_{aug})$, i.e., the $R({T_{\bar o}})$. 
Let $ {\begin{bmatrix}
        {{0_{n \times q}}}&{{I_{q \times q}}} 
\end{bmatrix}} = { {\begin{bmatrix}
            {r_1^ \top }&{r_2^ \top }& \ldots &{r_q^ \top } 
    \end{bmatrix}} ^ \top }$, then there exist $i \in \{1,\ldots,q\}$ and $\bar{T}_i \in \mathbb{R}^{1\times q}$ such that
\begin{equation}\label{eq_prove_nec1}
    \bar{T}_i T_{\bar{o}} = r_i^{\top},
\end{equation} where $\bar{T}_i$ is a linear map. 
This also implies that
\begin{equation}\label{eq_prove_necc1}
    r_i^{\top}\notin R(T_{{o}}).
\end{equation}
Then, by applying Corollary~1 in \cite{cong2021rethinking} to the system described by \eqref{eq_augaug_sys} and \eqref{eq_augaug_syso}, $\llbracket \tilde{\mathbf{z}}_k\rrbracket^- := \llbracket\tilde{\mathbf{z}}_k|u_{0:k-1},y_{0:k-1}\rrbracket$ and $\llbracket \tilde{\mathbf{z}}_k\rrbracket^+ := \llbracket\tilde{\mathbf{z}}_k|u_{0:k-1},y_{0:k}\rrbracket$ can be recursively derived by
\begin{align}
    \llbracket \tilde{\mathbf{z}}_k\rrbracket^- &= \tilde{A}\llbracket \tilde{\mathbf{z}}_{k-1}\rrbracket^+ \oplus {\tilde \Gamma }\{ {u_{k-1}}\}  \oplus {\tilde B}\llbracket{\Delta {{\mathbf{d}}_k}}\rrbracket, \label{eq_SMFobpre}\\
    \llbracket \tilde{\mathbf{z}}_k\rrbracket^+  &= {\mathscr{Z}}( \tilde{C},{y_{k}}) \cap \llbracket \tilde{\mathbf{z}}_k\rrbracket^-, \label{eq_SMFobup}	
\end{align}
where we define $\llbracket \tilde{\mathbf{z}}_{0}\rrbracket^- = \llbracket \tilde{\mathbf{z}}_{0}\rrbracket =T(\llbracket {\mathbf{z}}_{0}\rrbracket)$ with $\llbracket {\mathbf{z}}_{0}\rrbracket=\llbracket {{\mathbf{x}_0}}\rrbracket \times \llbracket {{\mathbf{ d}_0}}\rrbracket$ for consistency; ${\mathscr{Z}}(\tilde{C},{y_{k}}) = \{ {z_k}:{y_k} = \tilde{C}{z_k}\}$ is the affine space defined by ${\tilde{C}}$ and ${y_k}$.
Now, we prove
\begin{equation}
    D(\llbracket{\mathbf{d}}_k\rrbracket^+) = \infty, \quad k \in \mathbb{N}_0
\end{equation}
by mathematical induction.

\emph{Base case:} For $k=0$, we define $ {T_o}(\llbracket {\mathbf{z}}_{0}\rrbracket^-)=\mathcal S_0$ (with ${s_0} \in\mathcal S_0$). According to \eqref{eq_SMFobup}, one has
\begin{equation}\label{eq_basedcase_proof}
    \begin{split}
        \llbracket {{{{\mathbf{\tilde z}}}_0}} 
        \rrbracket^+ &= {\mathscr{Z}}(\tilde{C},{y_0}) \cap \llbracket \tilde{\mathbf{z}}_{0}\rrbracket^-
        \\&\supseteq({\mathscr{Z}}({{\tilde C}_o},{y_0}) \times {\mathbb{R}^{N_{\bar o}}}) \cap(\{ {s_0}\} \times {T_{\bar o}}\llbracket {{{\mathbf{z}}_0}|s_0} 
        \rrbracket^-)
        \\&= ({\mathscr{Z}}({{\tilde C}_o},{y_0}) \cap \{ {s_0}\}) \times {T_{\bar o}}\llbracket {{{{\mathbf{ z}}}_0}|{s_0}}\rrbracket^-
        \\&= \{ {s_0}\}  \times {T_{\bar o}}\llbracket {{{\mathbf{z}}_0}|{s_0}} 
        \rrbracket^-,
    \end{split} 
\end{equation}
where $N_{\bar o}$ is the dimension of unobservable states.
We define ${T^{ - 1}} =  {\begin{bmatrix}
        {{S_o}}&{{S_{\bar o}}} 
\end{bmatrix}} $. Then, we have
\begin{equation}\label{eq_basedcase_i}
    T{T^{ - 1}} = {\begin{bmatrix}
            {{T_o}{S_o}}&{{T_o}{S_{\bar o}}} \\ 
            {{T_{\bar o}}{S_o}}&{{T_{\bar o}}{S_{\bar o}}} 
    \end{bmatrix}} = I.
\end{equation}
From \eqref{eq_basedcase_proof}, one has
\begin{equation}\label{eq_basedcase_proof2}
    \begin{split}
        \llbracket {{{\mathbf{d}}_0}} 
        \rrbracket^+ &\mathop=\limits^{\eqref{eq_transform_P}}  {\begin{bmatrix}
                {{0_{q \times n}}}&{{I_{q \times q}}} 
        \end{bmatrix}}\llbracket {{{\mathbf{z}}_0}} 
        \rrbracket^+ \hfill 
        \\&= {\begin{bmatrix}
                {{0_{q \times n}}}&{{I_{q \times q}}} 
        \end{bmatrix}}{T^{ - 1}}\llbracket {{{\tilde{\mathbf{z}}}_0}} 
        \rrbracket^+ \hfill 
        \\&\mathop\supseteq\limits^{\eqref{eq_basedcase_proof}}  {\begin{bmatrix}
                {{0_{q \times n}}}&{{I_{q \times q}}} 
        \end{bmatrix}}  {\begin{bmatrix}
                {{S_o}}&{{S_{\bar o}}} 
        \end{bmatrix}} (\{ {s_0}\}  \times {T_{\bar o}}\llbracket {{{\mathbf{z}}_0}|{s_0}} 
        \rrbracket^-) \hfill 
        \\&=  {\begin{bmatrix}
                {{0_{q \times n}}}&{{I_{q \times q}}} 
        \end{bmatrix}}({S_o}\{ {s_0}\} \oplus  {S_{\bar o}}{T_{\bar o}}\llbracket {{{\mathbf{z}}_0}|{s_0}} 
        \rrbracket^-).\hfill 
    \end{split} 
\end{equation}
Let $\llbracket {{\mathbf{d}}_k^i} 
\rrbracket^+ = e_i^ \top \llbracket {{{\mathbf{d}}_k}} 
\rrbracket^+$, with $e_i^ \top$ satisfying ${e_i^ \top}
{\begin{bmatrix}
        {{0_{q \times n}}}&{{I_{q \times q}}} 
\end{bmatrix}}  = r_i^ \top$.
Then, from \eqref{eq_basedcase_i} and \eqref{eq_basedcase_proof2}, there exist $i \in \{1,\ldots,q\}$ and $\bar{T}_i \in \mathbb{R}^{1\times q}$ such that
\begin{equation}\label{eq_basedproof_potr}
    \begin{split}
        \llbracket {{\mathbf{d}}_0^i} 
        \rrbracket^+ &= e_i^ \top \llbracket {{\mathbf{d}}_0}\rrbracket^+  \\
        &\mathop\supseteq\limits^{\eqref{eq_basedcase_proof2}} {e_i^ \top}{\begin{bmatrix}
                {{0_{q \times n}}}&{{I_{q \times q}}} 
        \end{bmatrix}}({S_o}\{ {s_0}\}  \oplus {S_{\bar o}}{T_{\bar o}}\llbracket {{{\mathbf{z}}_0}|{s_0}} \rrbracket^-)  \\
        &= r_i^ \top ({S_o}\{ {s_0}\}  \oplus {S_{\bar o}}{T_{\bar o}}\llbracket {{{\mathbf{z}}_0}|{s_0}} \rrbracket^-) \\
        &\mathop=\limits^{\eqref{eq_prove_nec1}} \bar{T}_i T_{\bar{o}} ({S_o}\{ {s_0}\}  \oplus {S_{\bar o}}{T_{\bar o}}\llbracket {{{\mathbf{z}}_0}|{s_0}} \rrbracket^-) \\
        &\mathop=\limits^{\eqref{eq_basedcase_i}} r_i^ \top \llbracket {{{\mathbf{z}}_0}|s_0} \rrbracket^-
        \mathop=\limits^{(a)}\llbracket {{\mathbf{d}}_0^i} \rrbracket, \\ 
    \end{split}
\end{equation}
where $(a)$ follows from $r_i^ \top \llbracket {{{\mathbf{z}}_0}|s_0} \rrbracket^- = \llbracket {{\mathbf{d}}_0^i} | s_0\rrbracket$ and the fact that ${s_0}$ is unrelated to $\llbracket {{\mathbf{d}}_0^i} \rrbracket$ [see \eqref{eq_prove_necc1}], i.e., $\llbracket {{\mathbf{d}}_0^i|s_0} 
\rrbracket = \llbracket {{\mathbf{d}}_0^i} \rrbracket$.
Finally, by applying \lemref{lm_unboundedcombine}, it can be concluded that $D(\llbracket {{\mathbf{d}}_0^i} \rrbracket^+)=\infty$.
Therefore, $\llbracket {{{\mathbf{d}}_0}} 
\rrbracket^+$ is unbounded. 

\emph{Inductive step:} 
Assume $D(\llbracket {{\mathbf{d}}_{k'}^i}
\rrbracket^+) = \infty$ hold for any $k=k' \in {\mathbb{N}_0}$.
Then, according to \eqref{eq_dismodel} and \lemref{lm_unboundedcombine}, with $\llbracket\mathbf{d}_k\rrbracket^- := \llbracket\mathbf{d}_k|u_{0:k-1},y_{0:k-1}\rrbracket$, for $k=k'+1$, one has
\begin{equation}\label{key}
    D(\llbracket {{{\mathbf{d}}_{k' + 1}^i}} 
    \rrbracket^-) = D(\llbracket {{{\mathbf{d}}_{k'}^i}} 
    \rrbracket^+ \oplus \llbracket {\Delta {{\mathbf{d}}_{k'}^i}} 
    \rrbracket)=\infty,
\end{equation}
where $\Delta {{\mathbf{d}}_{k'}^i}$ stands for the $i$th entry of $\Delta {{\mathbf{d}}_{k'}}$.
Besides, similarly to \eqref{eq_basedproof_potr}, one has 
\begin{equation}\label{key}
    \llbracket {{\mathbf{d}}_{k' + 1}^i} 
    \rrbracket^+ \supseteq \llbracket {{\mathbf{d}}_{k' + 1}^i} 
    \rrbracket^-,
\end{equation}
which gives that $D(\llbracket {{{\mathbf{d}}_{k'+1}}} 
\rrbracket^-)=\infty$.

From the mathematical induction, we have $D({\llbracket {{\mathbf{d}}_{k}}
\rrbracket}^-)=\infty$ (w.r.t. $k \in {\mathbb{N}_0}$).
Thus, a bounded DO does not exist.\hfill$\blacksquare$

\section{Proof of \thmref{thm_OIT_bounded}}\label{apx_thm2}
From \eqref{eq_thm2pf_transform1}, if \eqref{eq_thm2} is satisfied, there exists a linear map $P$ such that the posterior range of observable subsystem states $\llbracket {{\mathbf{\tilde z}}_k^o} 
\rrbracket^+$ and actual disturbance range $\llbracket {{{\mathbf{d}}_k}} \rrbracket^+$ satisfy
\begin{equation}\label{eq_rtansformconclusion}
    P\llbracket {{\mathbf{\tilde z}}_k^o} 
    \rrbracket^+ = \llbracket {{{\mathbf{d}}_k}} 
    \rrbracket^+.
\end{equation}
Thus, according to Corollary~1 in \cite{cong2021rethinking}, we establish the bounded DO based on the SMF framework as follows:
\begin{align}
    \llbracket {{\mathbf{\tilde z}}_{k}^o} \rrbracket^- &= {{{\tilde A}_o}}\llbracket {{\mathbf{\tilde z}}_{k-1}^o} \rrbracket^+ \oplus {{{\tilde {\Gamma}}_o}}\{ {u_{k-1}}\}  \oplus {{\tilde {B}}_o} \llbracket{\Delta {{\mathbf{d}}_k}}\rrbracket, \label{eq_SMFobprenew}\\
    \llbracket  {{\mathbf{\tilde z}}_{k}^o} 
    \rrbracket^+  &= {\mathscr{Z}}( {{{\tilde C}_o}},{y_{k}}) \cap \llbracket {{\mathbf{\tilde z}}_{k}^o} \rrbracket^-, \label{eq_SMFobupnew}\\
    \llbracket {{{\mathbf{ d}}}_k}
    \rrbracket^+ &= P\llbracket {{\mathbf{\tilde z}}_{k}^o} 
    \rrbracket^+,\label{eq_SMFDoobnew}	
\end{align}
where ${{{\tilde A}_o}}\in {\mathbb{R}^{N_o \times N_o}}$, ${{{\tilde B}_o}}\in {\mathbb{R}^{N_o \times q}}$, ${{{\tilde \Gamma}_o}}\in {\mathbb{R}^{N_o \times p}}$ and ${{{\tilde C}_o}}\in {\mathbb{R}^{l \times N_o}}$, are defined by \eqref{eq_augaug_sys}.
The proposed algorithm is designed based on the SMF framework \eqref{eq_SMFobprenew}-\eqref{eq_SMFDoobnew}.
By outer bounding $\llbracket {\Delta {{\mathbf{d}}_k}}
\rrbracket$ with CZs, we can express the recursive step \eqref{eq_SMFobprenew}-\eqref{eq_SMFDoobnew} with CZ operation in \cite{Scott2016}, which gives \eqref{eq_CZ_predict} and \eqref{eq_CZ_update}, where the prior range and posterior range are expressed by $\mathcal{Z}_{k}^-$ and $\mathcal{Z}_{k}$ respectively.       
Then, by recursively using \eqref{eq_CZ_predict} and \eqref{eq_CZ_update}, we can obtain \eqref{eq_CZPARA}.
With the preparation above, we introduce a two-step proof of the boundedness of \algref{alg:thm}. 
In step 1, we prove $\mathcal{Z}_k^d \supseteq \llbracket {{{\mathbf{d}}_k}} \rrbracket^+$ $\forall k \in {\mathbb{N}_0}$. 
In step 2, we prove the boundedness of $\mathcal{Z}_k^d$.

\emph{Step~1 (Outer bound):}
For $k=0$, according to \eqref{eq_SMFobupnew}, we have
\begin{equation}
    \begin{split}
        \mathcal{Z}_0^d &= P{\mathcal{Z}_0} = P[\mathscr{Z}({{\tilde C}_o},{y_k}) \cap \mathcal{Z}_0^ - ]
        \\& \mathop  \supseteq \limits^{(a)} P[\mathscr{Z}({{\tilde C}_o},{y_k}) \cap \llbracket {{\mathbf{\tilde z}}_0^o}\rrbracket^-]\supseteq \llbracket  {{{\mathbf{d}}_0}}\rrbracket^+, 
    \end{split} 
\end{equation}
where $(a)$ follows from line 2 in \algref{alg:thm}.
For $k=1$, we have
\begin{equation}
    \begin{split}
        &\mathcal{Z}_{1}^d = P{\mathcal{Z}_{1}} = P[\mathscr{Z}({{\tilde C}_o},{y_{1}}) \cap \mathcal{Z}_{1}^ - ]
        \\&= P[\mathscr{Z}({{\tilde C}_o},{y_1}) \cap ({{\tilde A}_o}{\mathcal{Z}_{0}} \oplus {{\tilde \Gamma}_o}\{ {u_{0}}\}  \oplus \mathcal{Z}_{0}^\Delta )]
        \\&\supseteq P[\mathscr{Z}({{\tilde C}_o},{y_1}) \cap ({{\tilde A}_o}{\llbracket {{{\mathbf{\tilde z}}_{0}^o}} \rrbracket^+} \oplus {{\tilde \Gamma}_o}\{ {u_{0}}\}  \oplus {{\tilde B}_o}\llbracket {{{\Delta\mathbf{d}}_{0}}} 
        \rrbracket)]
        \\&\supseteq \llbracket {{{\mathbf{d}}_{1}}} 
        \rrbracket^+. 
    \end{split}
\end{equation}
Proceeding forward, we have $\mathcal{Z}_k^d \supseteq \llbracket {{{\mathbf{d}}_k}} \rrbracket^+$ $\forall k \in {\mathbb{N}_0}$ and $k < \delta$.

When $k = \delta$, we introduce the definition of filtering map as follows.
\begin{definition}[Filtering map]\label{def_filteringmap}
    At time $k\in {\mathbb{N}_0}$, the prediction-update map under system \eqref{eq_SMFobprenew} and \eqref{eq_SMFobupnew} is 
    \begin{equation}
        {f_k}:(\mathcal{S},\mathcal{M}_k) \mapsto \mathscr{Z}({{\tilde C}_o},{y_k}) \cap ({{\tilde A}_o}\mathcal{S} + {{\tilde \Gamma }_o}{u_{k - 1}} + {{\tilde B}_o}\mathcal{M}_k),
    \end{equation}
    where $\mathcal{S} \in {\mathbb{R}^n}$ and $\mathcal{M}_k \in {\mathbb{R}^q}$. The filtering map is ${F_{k,i}}$ that
    \begin{equation*}
        {F_{k,i}}(\mathcal{S},{\mathcal{M}_i}, \ldots ,{\mathcal{M}_k}) = {{f_k} \circ(  \cdots  \circ ({f_{i + 1}}(\mathcal{S},{\mathcal{M}_i}),\ldots),\mathcal{M}_k}).
    \end{equation*}
\end{definition}
According to \algref{alg:thm}, since \eqref{eq_CZPARA} is derived by recursively using \eqref{eq_CZ_predict} and \eqref{eq_CZ_update}, we can explain \eqref{eq_CZPARA} with a filtering map, which gives
\begin{equation}
    \begin{split}
        {\mathcal{Z}_k^d} &= P{F_{k,k - \delta }}[\overline {{\text{IH}}} ({\mathcal{Z}_{k - \delta }}),\mathcal{Z}_{k - \delta }^\Delta , \ldots ,\mathcal{Z}_k^\Delta ] \\&
        \supseteq P{F_{k,k - \delta }}(\llbracket {{\mathbf{\tilde z}}_{k - \delta }^o} 
        \rrbracket^+,\llbracket {\Delta {{\mathbf{d}}_{k - \delta }}} 
        \rrbracket, \ldots ,\llbracket {\Delta {{\mathbf{d}}_k}} 
        \rrbracket) \\&
        \supseteq P{F_{k,0 }}(\llbracket {{\mathbf{\tilde z}}_{0 }^o} 
        \rrbracket^+,\llbracket {\Delta {{\mathbf{d}}_{0 }}} 
        \rrbracket, \ldots ,\llbracket {\Delta {{\mathbf{d}}_k}} 
        \rrbracket) \\&
        = P\llbracket {{\mathbf{\tilde z}}_k^o} 
        \rrbracket = \llbracket {{{\mathbf{d}}_k}} \rrbracket^+.
    \end{split}
\end{equation}
Proceeding forward, from the mathematical induction, we have $\mathcal{Z}_k^d \supseteq \llbracket {{{\mathbf{d}}_k}} \rrbracket^+$ $\forall k \in {\mathbb{N}_0}$ and $k>\delta$.

Therefore, we can conclude that $\mathcal{Z}_k^d \supseteq \llbracket {{{\mathbf{d}}_k}} \rrbracket^+$ $\forall k \in {\mathbb{N}_0}$.
 
\emph{Step~2 (Boundedness):}
From our previous work \cite{cong2022stability}, we know that $\mathcal{Z}_{k}$ derived from \algref{alg:thm} is uniformly bounded w.r.t $k \in {\mathbb{N}_0}$.
Then, one has
\begin{equation}\label{eq_algor_suf_stability}
    \begin{split}
        &{\varlimsup_{k\to\infty}\sup _{{{\hat d}_k}\in \mathcal{Z}_{k}^{\Delta},{d_k} \in \llbracket {{{\mathbf{d}}_k}} 
                \rrbracket^+}}\left\| {{{\hat d}_k} - {d_k}} \right\|
        \\& = \mathop {\overline {\lim } }\limits_{k \to \infty } \mathop {\sup }\limits_{\hat {\tilde z}_k^o \in {\mathcal{Z}_k},\tilde z_k^o \in \llbracket {{\mathbf{\tilde z}}_k^o} \rrbracket^+} \left\| {P\hat {\tilde z}_k^o - P\tilde z_k^o} \right\|
        \\&  \leqslant  \mathop {\overline {\lim } }\limits_{k \to \infty } \mathop {\sup }\limits_{\hat {\tilde z}_k^o, \tilde z_k^o \in {\mathcal{Z}_k}} \left\| {P\hat {\tilde z}_k^o - P\tilde z_k^o} \right\|
        < \infty.
    \end{split}
\end{equation}
Therefore, $\hat{d}_k$ is bounded w.r.t. $k \in {\mathbb{N}_0}$.\hfill$\blacksquare$

\section{Proof of \thmref{thm:Worst-Case Optimality}}\label{apx:Proof of thm:Worst-Case Optimality}

When \eqref{eq_optimal_re1} is satisfied, we have 
\begin{equation*}
    \begin{split}
        \mathcal{Z}_{0}^d &= P[\mathscr{Z}({{\tilde C}_o},{y_0}) \cap \llbracket \tilde{\mathbf{z}}_0^o\rrbracket]= \llbracket \mathbf{d}_0|y_0\rrbracket = \llbracket  {{{\mathbf{d}}_0}}\rrbracket^+,\\
        \mathcal{Z}_{1}^d &= P[\mathscr{Z}({{\tilde C}_o},{y_1}) \cap ({{\tilde A}_o}{\llbracket \tilde{\mathbf{z}}_0^o|y_0 \rrbracket} \oplus {{\tilde \Gamma}_o}\{ {u_{0}}\}  \oplus {{\tilde B}_o}\llbracket {{{\Delta\mathbf{d}}_{0}}} 
            \rrbracket)]
        \\&=\llbracket  {{{\mathbf{d}}_1}}\rrbracket^+,
    \end{split}
\end{equation*}
at $k=0$ and $k=1$, repectively.
Proceeding forward, we have $\mathcal{Z}_k^d=\llbracket{{{\mathbf{d}}_k}}\rrbracket^+$ when $k<\delta$. 
According to \algref{alg:thm}, we choose the centroid of the interval hull of $\mathcal{Z}_{k}^d$ as the estimate (i.e., $\hat d_k = {\text{center}}(\overline {{\text{IH}}}( \mathcal{Z}_{k}^d ) )$),
which gives
\begin{equation}\label{eq_CZ_optimap2}
    \sup_{d_k \in \llbracket\mathbf{d}_k\rrbracket^+} \|\hat d_k - d_k\| \leq \sup_{d_k \in \llbracket\mathbf{d}_k\rrbracket^+} \|\hat{d}_k' - d_k\|				
\end{equation}
for any ${{\hat d'}_k} \in {\mathbb{R}_q}$.
Thus,~\eqref{eqn:Optimal Estimate} is satisfied.\hfill$\blacksquare$

\bibliographystyle{IEEEtran}
\bibliography{ww}

\begin{thebibliography}{10}
\providecommand{\url}[1]{#1}
\csname url@samestyle\endcsname
\providecommand{\newblock}{\relax}
\providecommand{\bibinfo}[2]{#2}
\providecommand{\BIBentrySTDinterwordspacing}{\spaceskip=0pt\relax}
\providecommand{\BIBentryALTinterwordstretchfactor}{4}
\providecommand{\BIBentryALTinterwordspacing}{\spaceskip=\fontdimen2\font plus
\BIBentryALTinterwordstretchfactor\fontdimen3\font minus
  \fontdimen4\font\relax}
\providecommand{\BIBforeignlanguage}[2]{{%
\expandafter\ifx\csname l@#1\endcsname\relax
\typeout{** WARNING: IEEEtran.bst: No hyphenation pattern has been}%
\typeout{** loaded for the language `#1'. Using the pattern for}%
\typeout{** the default language instead.}%
\else
\language=\csname l@#1\endcsname
\fi
#2}}
\providecommand{\BIBdecl}{\relax}
\BIBdecl

\bibitem{chen2015disturbance}
W.-H. Chen, J.~Yang, L.~Guo, and S.~Li, ``Disturbance-observer-based control
  and related methods—an overview,'' \emph{IEEE Trans. Ind. Electron.},
  vol.~63, no.~2, pp. 1083--1095, Feb 2016.

\bibitem{Wang2016}
X.~Wang, S.~Li, X.~Yu, and J.~Yang, ``Distributed active anti-disturbance
  consensus for leader-follower higher-order multi-agent systems with
  mismatched disturbances,'' \emph{IEEE Trans. Autom. Control}, vol.~62,
  no.~11, pp. 5795--5801, Nov 2017.

\bibitem{8809884}
K.~Zhang, B.~Jiang, M.~Chen, and X.-G. Yan, ``Distributed fault estimation and
  fault-tolerant control of interconnected systems,'' \emph{IEEE Trans.
  Cybern.}, vol.~51, no.~3, pp. 1230--1240, Mar 2021.

\bibitem{kurkccu2018disturbance}
B.~K{\"u}rk{\c{c}}{\"u}, C.~Kasnako{\u{g}}lu, and M.~{\"O}. Efe,
  ``Disturbance/uncertainty estimator based integral sliding-mode control,''
  \emph{IEEE Trans. Autom. Control}, vol.~63, no.~11, pp. 3940--3947, Nov 2018.

\bibitem{10124355}
H.~Liu, Y.~Li, Q.-L. Han, T.~Raïssi, and T.~Chai, ``Secure estimation, attack
  isolation and reconstruction based on zonotopic unknown input observer,''
  \emph{IEEE Trans. Autom. Control}, pp. 1--13, May 2023.

\bibitem{lan2020asymptotic}
J.~Lan, ``Asymptotic estimation of state and faults for linear systems with
  unknown perturbations,'' \emph{Automatica}, vol. 118, p. 108955, Aug 2020.

\bibitem{rabiee2019continuous}
H.~Rabiee, M.~Ataei, and M.~Ekramian, ``Continuous nonsingular terminal sliding
  mode control based on adaptive sliding mode disturbance observer for
  uncertain nonlinear systems,'' \emph{Automatica}, vol. 109, p. 108515, Nov
  2019.

\bibitem{chang2006applying}
J.-L. Chang, ``Applying discrete-time proportional integral observers for state
  and disturbance estimations,'' \emph{IEEE Trans. Autom. Control}, vol.~51,
  no.~5, pp. 814--818, May 2006.

\bibitem{Kim2013}
K.-S. Kim and K.-H. Rew, ``Reduced order disturbance observer for discrete-time
  linear systems,'' \emph{Automatica}, vol.~49, no.~4, pp. 968--975, Apr 2013.

\bibitem{Corless1998}
M.~Corless and J.~Tu, ``State and input estimation for a class of uncertain
  systems,'' \emph{Automatica}, vol.~34, no.~6, pp. 757--764, Jun 1998.

\bibitem{alenezi2021simultaneous}
B.~Alenezi, M.~Zhang, S.~Hui, and S.~H. {\.Z}ak, ``Simultaneous estimation of
  the state, unknown input, and output disturbance in discrete-time linear
  systems,'' \emph{IEEE Trans. Autom. Control}, vol.~66, no.~12, pp.
  6115--6122, Dec 2021.

\bibitem{9872111}
M.~Shen, T.~Zhang, J.~H. Park, Q.-G. Wang, and L.-W. Li, ``Iterative
  proportional-integral interval estimation of linear discrete-time systems,''
  \emph{IEEE Trans. Autom. Control}, vol.~68, no.~7, pp. 4249--4256, Jul 2023.

\bibitem{darouach2000existence}
M.~Darouach, ``Existence and design of functional observers for linear
  systems,'' \emph{IEEE Trans. Autom. Control}, vol.~45, no.~5, pp. 940--943,
  May 2000.

\bibitem{5457987}
K.-S. Kim, K.-H. Rew, and S.~Kim, ``Disturbance observer for estimating higher
  order disturbances in time series expansion,'' \emph{IEEE Trans. Autom.
  Control}, vol.~55, no.~8, pp. 1905--1911, Aug 2010.

\bibitem{Su2018}
J.~Su and W.-H. Chen, ``Further results on “reduced order disturbance
  observer for discrete-time linear systems”,'' \emph{Automatica}, vol.~93,
  pp. 550--553, Jul 2018.

\bibitem{8732471}
I.~Sakhraoui, B.~Trajin, and F.~Rotella, ``Design procedure for linear unknown
  input functional observers,'' \emph{IEEE Trans. Autom. Control}, vol.~65,
  no.~2, pp. 831--838, Feb 2020.

\bibitem{6415998}
G.~N. Nair, ``A nonstochastic information theory for communication and state
  estimation,'' \emph{IEEE Trans. Autom. Control}, vol.~58, no.~6, pp.
  1497--1510, Jun 2013.

\bibitem{cong2021rethinking}
Y.~Cong, X.~Wang, and X.~Zhou, ``Rethinking the mathematical framework and
  optimality of set-membership filtering,'' \emph{IEEE Transactions on
  Automatic Control}, vol.~67, no.~5, pp. 2544--2551, 2021.

\bibitem{cong2022stability}
------, ``Stability of linear set-membership filters,''
  \emph{arXiv:2203.13966}, 2022.

\bibitem{Scott2016}
J.~K. Scott, D.~M. Raimondo, G.~R. Marseglia, and R.~D. Braatz, ``Constrained
  zonotopes: A new tool for set-based estimation and fault detection,''
  \emph{Automatica}, vol.~69, pp. 126--136, Jul 2016.

\end{thebibliography}
\end{document}